\documentclass[aps,pre,twocolumn,showpacs,amssymb,groupedaddress]{revtex4}
\usepackage{graphicx}
\usepackage{subfigure}

\begin{document}

\title{Maximum and Minimum Stable Random Packings of Platonic Solids}
\author{Jessica Baker and Arshad Kudrolli}

\affiliation{Department of Physics, Clark University, Worcester, MA
01610}

\date{\today} 
\begin{abstract}
Motivated by the relation between particle shape and packing, we measure the 
volume fraction $\phi$ occupied by the Platonic solids which are a class of  polyhedron with congruent sides, vertices and dihedral angles. Tetrahedron, cube, octahedron, dodecahedron, and icosahedron shaped plastic dice were fluidized or mechanically vibrated to find stable random loose packing $\phi _{rlp} = 0.51, 0.54, 0.52, 0.51, 0.50$ and densest packing $\phi_{rcp} = 0.64, 0.67, 0.64, 0.63, 0.59$, respectively with standard deviation $\simeq \pm 0.01$. We find that $\phi$ obtained by all protocols peak at the cube, which is the only Platonic solid that can tessellate space, and then monotonically decrease with number of sides. This overall trend is similar but systematically lower than the maximum $\phi$ reported for frictionless Platonic solids, and below $\phi_{rlp}$ of spheres for the loose packings. Experiments with ceramic tetrahedron were also conducted, and higher friction  was observed to lead to lower $\phi$. 
\end{abstract}


\maketitle
\section{Introduction}
The packing of objects has long fascinated physicists, mathematicians, and the curious. While the centuries old Kepler's conjecture that the maximum packing of spheres is $\pi/\sqrt{18} \simeq 0.74$ has been finally proven~\cite{hales}, spheres tossed randomly into a jar and shaken, do not reach such high packing fractions, unless special protocols are used~\cite{pouliquen}. A random closed packed (RCP) volume fraction $\phi_{rcp} = 0.6366$ was found by mechanically vibrating a set of steel ball bearings~\cite{scott}. 
In fact, work in the last decade has shown that random packing itself is not unique, and there is a range of packing fractions which can be obtained for spherical particles with random order. These values are bounded at the upper end by $\phi_{mrj} = 0.64$~\cite{torquato}, the so called maximally random close packed state, and more tentatively at the lower end by random loose packing at $\phi_{rlp} = 0.55$~\cite{onoda,rlp}.

\begin{figure}
\begin{center}
\includegraphics[width=7cm]{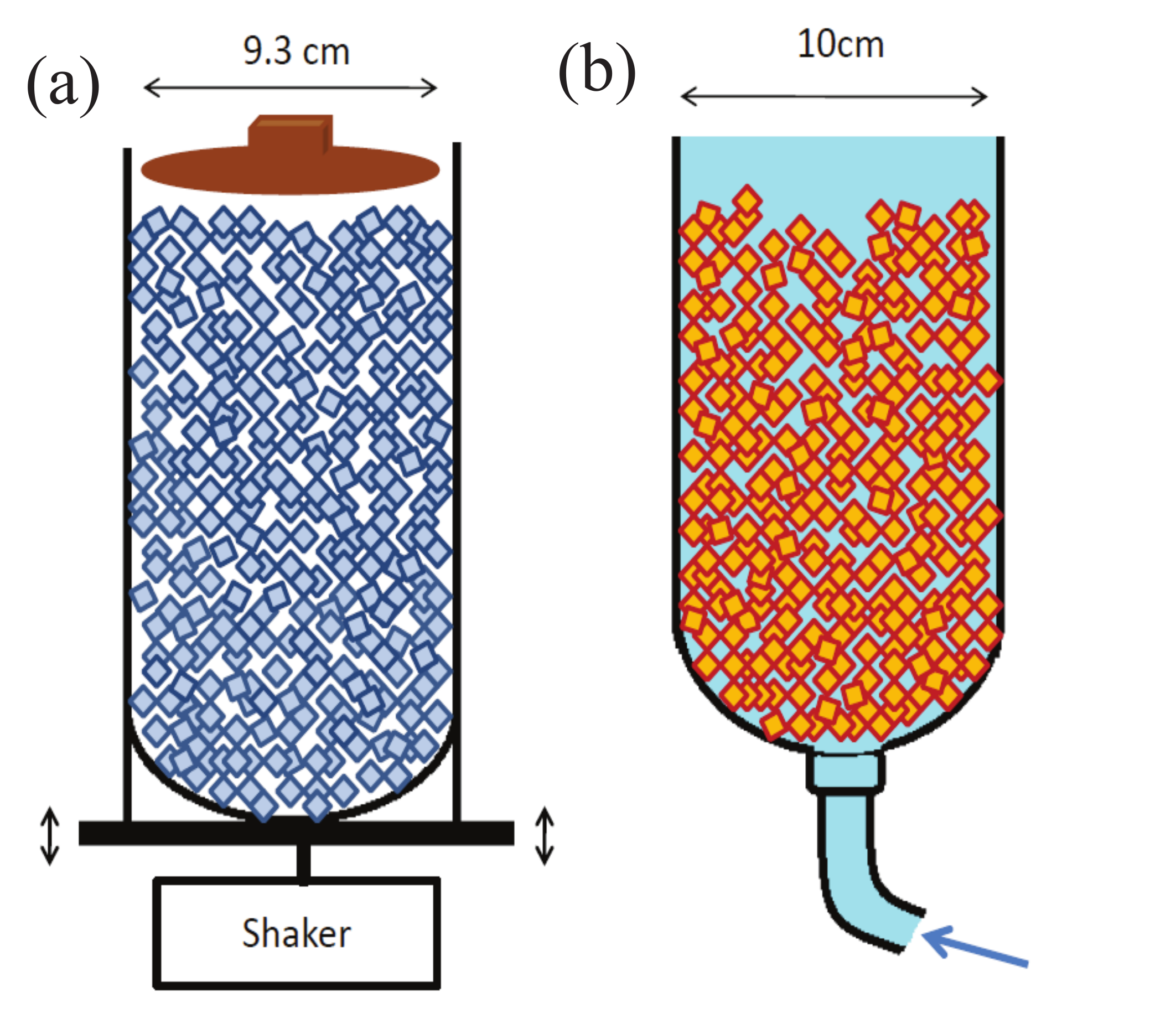}
\end{center}
\caption{(Color online) (a) A schematic diagram of the container which is vibrated vertically with an electromagnetic shaker to obtain the mechanically shaken packings. (b) A schematic diagram of the container used to obtain loose packings using the  fluidization protocol. The liquid is first injected from below and the packing height measured after flow rate is turned off and the liquid drained. \label{protocol}}
\end{figure}
In contrast, much less is known for non-spherical particles. It was only shown a few years ago that randomly packed prolate and oblate objects pack denser than spherical particles~\cite{mandm}. The highest Bravais lattice packing of Platonic solids are also considered to be the highest packings except in case of the tetrahedrons~\cite{betke,torquato2}. In the case of tetrahedrons, disordered wagon-wheel packings were initially found to pack even higher. The maximum packing of tetrahedrons has since been improved upon in rapid succession with different approaches~\cite{chen,torquato3}, and the current highest packing for tetrahedrons stands at 0.856347.. corresponding to dimer packings of regular tetrahedrons~\cite{kallus10,torquato10,chen10}. As noted for spheres, maximum packing tend to be larger than polyhedral packings which may be disordered or ordered when brought together randomly. For example, quasi-crystals were observed with Monte Carlo simulation of tetrahedrons by Haji-Akbari et al~\cite{haji}. Experiments on random packed tetrahedronal dice have been reported recently in Ref.~\cite{jaoshvili}. Volume fractions were said to be $0.76\pm.02$ if the observed packings were extrapolated to infinite systems, but the protocol by which the packings were prepared was not clear.

\begin{table*}[t]
\centering
\begin{tabular}{ccccccc}
\hline

 & \includegraphics[height=2cm]{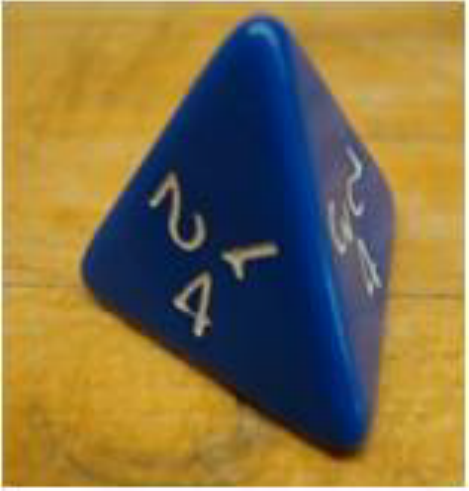} & \includegraphics[height=2cm]{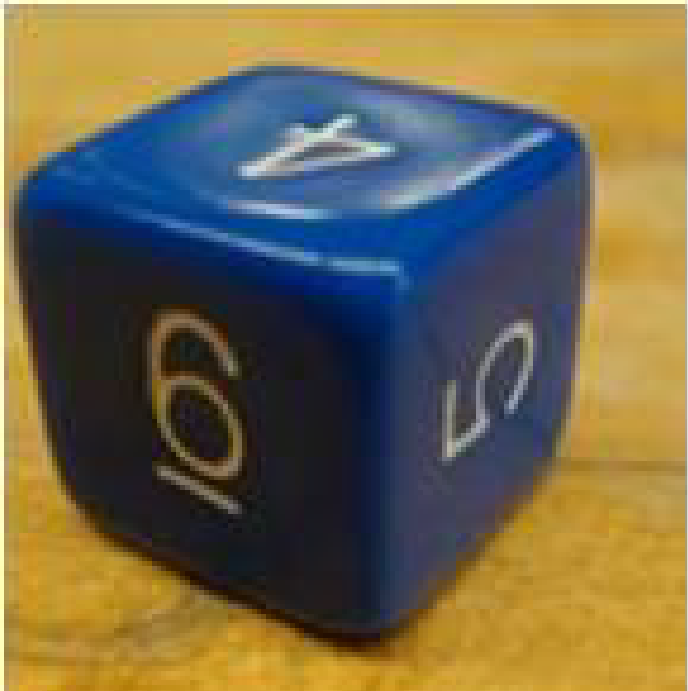} & \includegraphics[height=2cm]{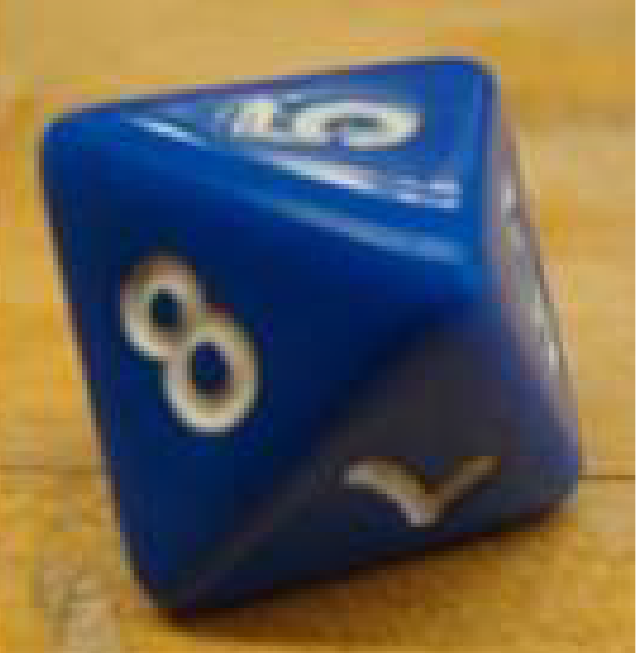} & \includegraphics[height=2cm]{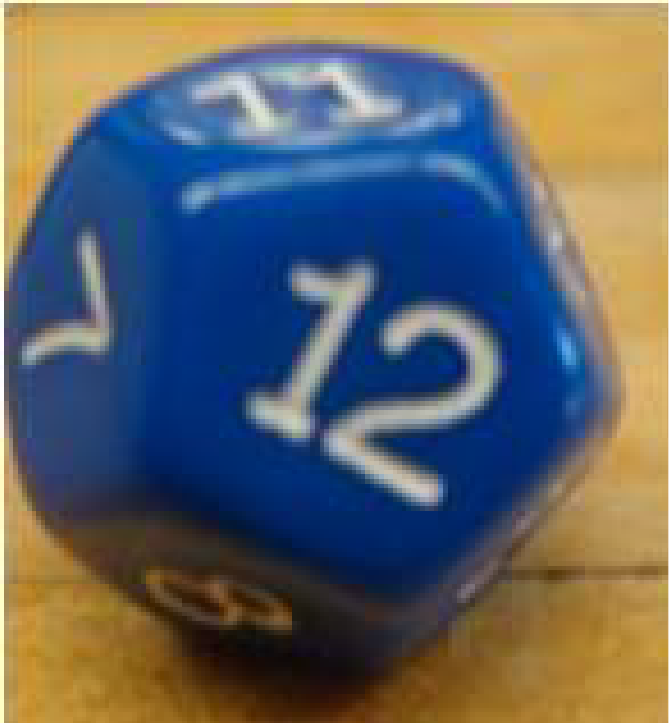} & \includegraphics[height=2cm]{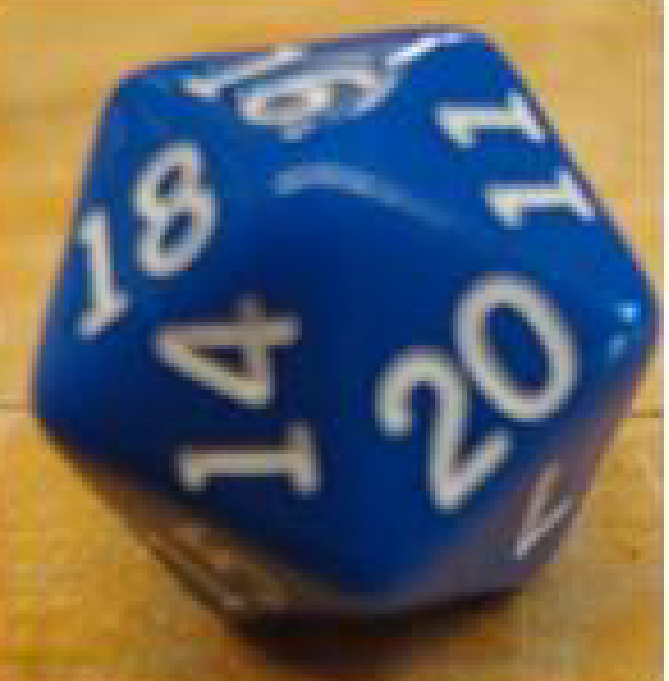} & \includegraphics[height=2cm]{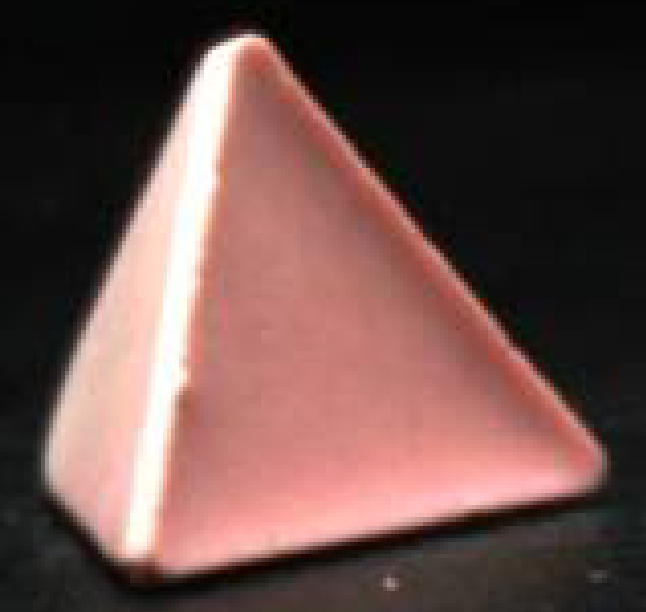}\\
\hline
\hline
Name & Tetrahedron & Cube & Octahedron & Dodecahedron & Icosahedron & Tetrahedron\\
Material & Plastic & Plastic & Plastic & Plastic & Plastic & Ceramic\\
Number of Faces & 4 & 6 & 8 & 12 & 20 & 4\\
Dihedral Angle (radian) & $ \cos^{-1} (\frac{1}{3})$ & $ \frac{\pi}{2} $ & $ \cos^{-1}(\frac{-1}{3})$ & $ \cos^{-1}(\frac{-\sqrt{5}}{5}) $ & $\cos^{-1}(\frac{-\sqrt{5}}{3}) $ & $ \cos^{-1} (\frac{1}{3})$ \\
Solid Angle (steradian) &$ \cos^{-1}(\frac{23}{27})$ & $ \frac{\pi}{2}$ & $ 4\sin^{-1}(\frac{1}{3}) $  &$ \pi - \tan^{-1}(\frac{2}{11}) $  &$ 2\pi - 5\sin^{-1}(\frac{2}{3}) $  & $ \cos^{-1}(\frac{23}{27}) $\\
$V_{solid}$  (cm$^3$) & $1.56$ & $4.0$ & $2.4$ & $4.0$ & $3.5$ & $1.7$\\
\hline
\end{tabular}
\caption{(Color online) The properties of the Platonic solids studied. The volume for each kind of solid $V_{solid}$ were found by averaging over 10 trials with water displacement measurements. }
    \label{dice}
\end{table*}

In this paper, we experimentally investigate the packings obtained with the Platonic solids using various experimental preparation protocols including sequential addition (unshaken), sequential addition with hand shaking, mechanical vibration, and fluidization. A question we also seek to address is if the packing fraction for faceted particles approaches that for spheres from above or below in the limit of large number of faces. We also test the effect of the number of particles and the friction between them on the packing. A further motivation for our study comes from the fact that natural sand is often faceted  with sharp edges which can strongly influence their packing density~\cite{cho}. Platonic solids which are idealized faceted particles with congruent sides may be a better starting point compared with smooth spheres to understand  packing of rough particles. 

\section{Experimental method}
\subsection{Materials}
The Platonic solids are a class of convex polyhedrons with faces of congruent polygons and the same number of faces meeting at each vertex. These conditions lead to congruent faces, dihedral angles and solid angles (Table~\ref{dice}). There are only 5 shapes that belong to this category:  the 4 sided tetrahedron, 6 sided cube, 8 sided octahedron, 12 sided icosahedrons and 20 sided dodecahedron.  The actual particles studied in our experiment are plastic dice (density $\rho = 1.16$\,g/cm$^3$) which have slightly rounded edges with properties listed in Table~\ref{dice}. To estimate the actual volume fractions occupied by the solids used, we measured the volume using water displacement technique. We thus assume that we are calculating the volume fraction of a Platonic solid which is in between the circumscribed and inscribed limit because of the rounded edges. To understand the effect of the rounding further, one has to obtain the distribution of contacts that involve vertices, edges, and the distribution of contact angles, which is beyond the scope of our technique. The predominance of these kind of contacts may indicate that we are systematically over estimating the volume fraction. On the other hand predominance of face-face contacts may lead to an underestimate. While it is possible some of these effects offset each other, it is difficult estimate the net error without detailed understanding on contacts. 

Ceramic tetrahedrons ($\rho = 1.63$\,g/cm$^3$) were also used to compare number of particles and effects of friction on packing fractions. The two kinds of materials have slightly different coefficient of friction ($\mu_{plastic} = 0.375$, $\mu_{ceramic} = 0.480$), which is measured using a tilted plane and finding the angle at which particles begin to slide past each other. Our system is athermal and therefore energy has to be supplied externally to rearrange the solids.  

\subsection{Packing preparation protocols}
\label{exptmethod}
We use four different random packing protocols to prepare the packing of Platonic solids in cylindrical containers with a semi-hemispherical bottom boundary to minimize surface area. We found this shape best suited to determine volume fractions accurately because even though a spherical container has a low surface to volume ratio, it is practically difficult to fill a particle under an overarching surface. 

\subsubsection{Sequential addition} In the first packing protocol, the particles are added sequentially at a random location from a height of about a few times the particle size. This ensured that the particles land in a stable configuration without significantly moving the particles which were already in place in the packing. 

A thin plate is placed on top after the packing is prepared and the average height is noted. Using this height, the total volume of the container $V_{container}$ is obtained to determine the volume fraction occupied by the solids $\phi = N V_{solid}/ V_{container}$, where $N$ is the number of particles added. This packing protocol was repeated 10 times for each kind of solid to determine the mean packing fraction and standard deviation. 

\subsubsection{Hand shaken} In the second method, the container is shaken by hand after two layers of particles are added to the container so that particles rattle and have an opportunity to rearrange. After all the particles are added, $\phi$ is determined (as above) from the measured height of the packing. As we will see this protocol leads to relatively higher $\phi$.   

\begin{figure*}
\begin{center}
\leavevmode
\includegraphics[width =16cm]{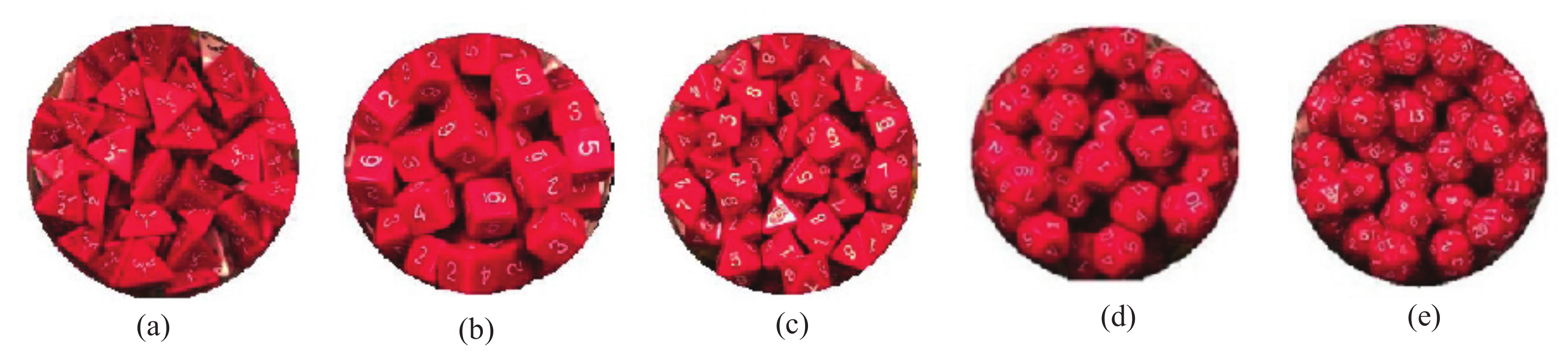}
\caption{(Color online) Images of typical disordered packings observed for (a) tetrahedrons, (b) cubes, (c) octahedrons, (d) dodecahedron, and (e) icosahedron. The images shown correspond to the top surface of the packing obtained after sequential addition of particles.}
\label{topimage}
\end{center}
\end{figure*}

\subsubsection{Mechanically shaken} Because it is impossible to shake the particles systematically by hand, we also built an experimental system in which the container is rigidly attached to an electromagnetic shaker (see schematic diagram shown in Fig.~\ref{protocol}(a)) similar to previous systems used to study random close packing of spheres~\cite{scott}. The shaker is connected to a function generator allowing us to apply a prescribed vibration frequency $f$ and acceleration strength $\Gamma$. After initially placing the particles randomly inside the container, and vibrating over various lengths of times, the height of the packing is recorded after the vibration is turned off to ensure that the packing obtained is stable. The volume fraction $\phi$ is then obtained as described in the unshaken case. This protocol gave rise to the highest packing fractions that we observed. 

We first performed measurement with experiments over 120 minutes of vibration, and found that the packing fractions increased rapidly initially by about 5\% in a few minutes and then did not vary significantly. Subsequently, we performed and measure $\phi$ after 10 minutes of applied vibration. While slow evolution of volume fraction over prolonged periods of vibrations 
has been observed with spherical particles~\cite{nowak98}, it appears that Platonic solids, 
which are faceted, get frustrated more quickly because they cannot rotate and roll 
as easily in place as spherical particles. We also tested the frequency dependence of obtained $\phi$ and found that a peak was observed at $f \sim 50$\,Hz. Because we are primarily interested in this method to obtain the maximum possible density using random agitations, we report here data for $f \sim 50$\,Hz and $5g$ where $g$ is the acceleration due to gravity. We present the frequency dependence of the observed packing fraction later in Sec.~\ref{freqdep}.

\subsubsection{Fluidization} In the final protocol used to prepare the packing, we first fill the particles inside a container with a hemispherical mesh at the bottom through which water can be injected with rates high enough to fully fluidize the plastic dice (see Fig.~\ref{protocol}(b)). A flow rate of $1.6 \times 10^4$ cm$^3$/min is applied for 3 minutes to fully agitate the system, and then the flow rate is slowly reduced to zero so that particles slowly settle layer by layer with low relative velocities. The water is then completely drained from the system and the height of the packing is measured to obtain $\phi$. This set up and method is similar to that used to obtain the limit of random loose packing in spheres~\cite{rlp}, and we also find the lowest $\phi$ for the Platonic solids among all protocols we attempted.

\section{Observed packings}
Figure~\ref{topimage} shows an image corresponding to typical packings observed for each of the Platonic solids. In this case, 200 particles were added inside a container with a 9.3 cm diameter using the sequential addition protocol and the image was taken of the top layer which is also similar to any intermediate layer. The particles appear to be all randomly located relative to each other. (We quantify the disorder in the packing later in Sec.~\ref{sec:order} using the variance of the orientation of the face of the polyhedron relative to the vertical axis.)  The packings prepared using the other protocols appear similarly random. 

The packing fractions for each of the Platonic solids obtained using the protocols described in Section~\ref{exptmethod} are listed in Table.~\ref{exp_phi}. 
As expected, the fluidized protocol creates the least dense packing which we consider as $\phi_{rlp}$ , and the mechanical shaker produces the most dense packing which we consider as $\phi_{rcp}$. Further we note that the sequential addition (unshaken) and hand shaken protocols produce intermediate packings. In all cases $\phi$ vary similarly and peak at the cube and then decrease with increasing number of sides.

\begin{table*}[t]
\centering
\begin{tabular}{ccccc}
\hline
Shape & $\phi_{\rm Sequential Addition}$ & $\phi_{\rm Hand Shaken}$ & $\phi_{\rm Mechanically Shaken}$ & $\phi_{\rm Fluidization}$ \\
\hline
\hline
Tetrahedon (Plastic) &  $ 0.54 \pm 0.01$ &  $ 0.62 \pm 0.02$ &  $ 0.64 \pm 0.01$ & $ 0.51 \pm 0.01$\\
Cube (Plastic) &  $ 0.57 \pm 0.01$ &  $ 0.66 \pm 0.02$ &  $ 0.67 \pm 0.02$ & $ 0.54 \pm 0.01$\\
Octahedron (Plastic) &  $ 0.57 \pm 0.01$ &  $ 0.62 \pm 0.01$ &  $ 0.64 \pm 0.01$ & $ 0.52 \pm 0.01$\\
Dodecahedron (Plastic) &  $ 0.56 \pm 0.01$ &  $ 0.60 \pm 0.01$ &  $ 0.63 \pm 0.01$ & $ 0.51 \pm 0.01$\\
Isocahedron (Plastic) &  $ 0.53 \pm 0.01$ &  $ 0.57 \pm 0.01$ &  $ 0.59 \pm 0.01$ & $ 0.50 \pm 0.01$\\
Tetrahedon (Ceramic) &  $ 0.48 \pm 0.02$ &  $ 0.59 \pm 0.01$ & - & - \\
\hline
\end{tabular}
\caption{The mean packing fractions ($ \pm$ one standard deviation) observed for each of the Platonic solids using the four protocols.}
    \label{exp_phi}
\end{table*}

\begin{figure}
\begin{center}
\includegraphics[scale = 0.4] {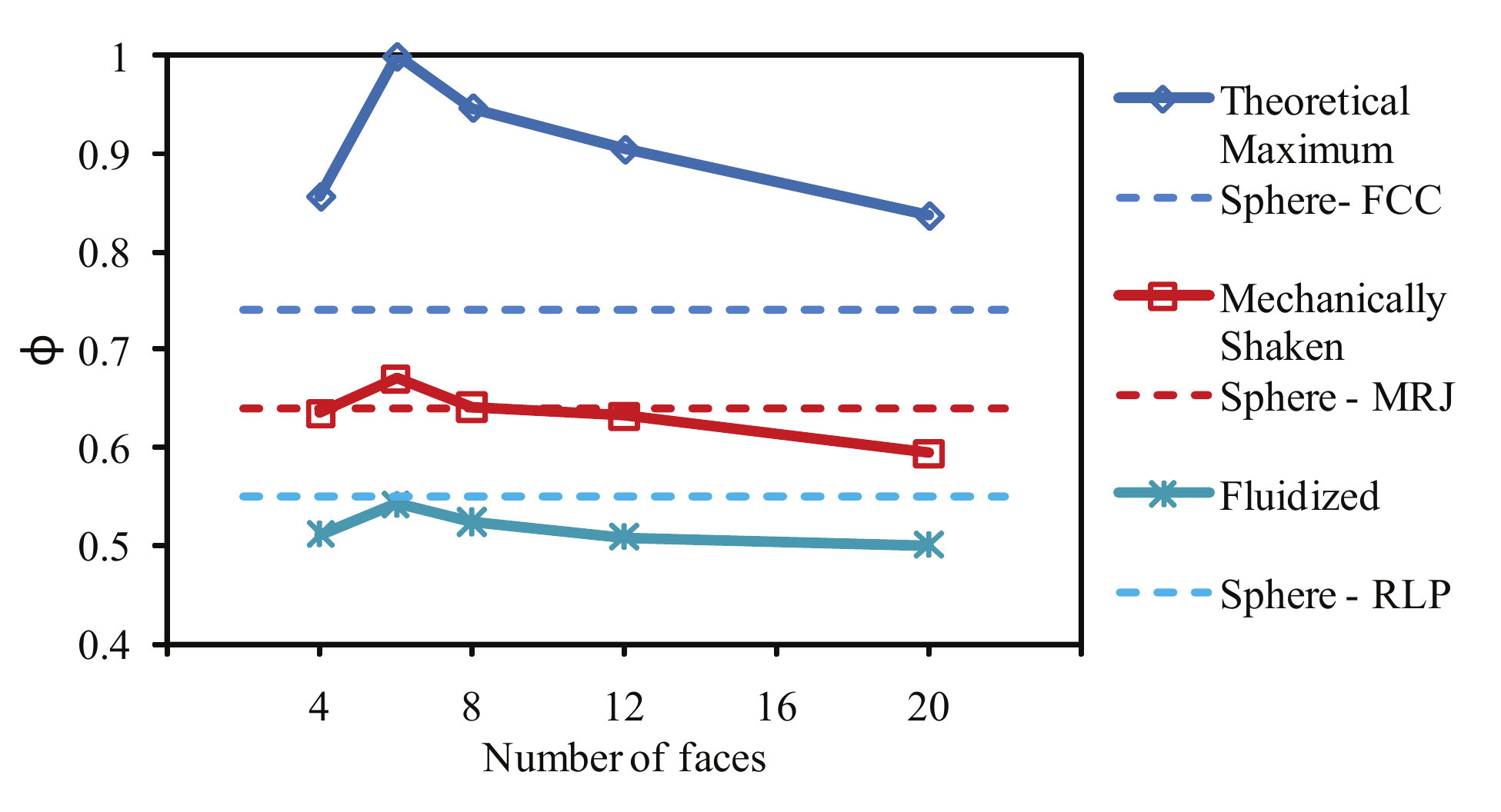}
\caption{(Color online) The packing fraction for the Platonic solids compared with the maximum known packing for that solid. The observed random packings follow the same trend as the maximum packings. The maximum value for tetrahedron is from Ref.~\cite{chen10}, and for the remaining Platonic solids are from Ref.~\cite{betke}. The maximum packing along with the RCP and RLP value for spheres is also plotted for comparison. }
\label{maxpacking}
\end{center}
\end{figure}

To understand the trends in the observed packings, we have plotted the mean values for each kind of solid and protocol in Fig.~\ref{maxpacking} along with the maximum known $\phi$ which have been reported~\cite{betke,chen10}. Interestingly,  we find that the data follows the same trend as the theoretical maximum $\phi$, but is systematically lower because of the disordered nature of their packing. While lower $\phi$ than maximum can be anticipated, it is somewhat unexpected that the trend for disordered packing reflects the maximum packing. 

\section{Supplementary measurements}

\subsection{Effect of Friction}
Because the particles we use have friction, and the presence of friction reduces the minimum number of particle contacts needed for static equilibrium from 3 to 2 in three dimensions, it is important to consider its effect on packing. In the case of spheres, friction is known to affect packing fractions especially at the loose packing limit~\cite{farrell}. Needing fewer neighbors implies the particles are further apart and therefore the packing can be less dense. It is possible for this reason that we observe a lower dense close packing value for tetrahedrons compared with simulations which found 0.6817 with frictionless tetrahedrons using a relaxation algorithm~\cite{shui08}. To explore this further, we compare packings of the plastic tetrahedrons with the more frictional ceramic tetrahedrons. We find that for both the unshaken and hand shaken protocols (see Table~\ref{exp_phi}), the larger the coefficient of friction, the lower the packing fraction with a 9\% and 5\% lower packing fraction for the ceramic case.

\subsection{System size}
\begin{figure}
\begin{center}
\includegraphics[scale = 0.4] {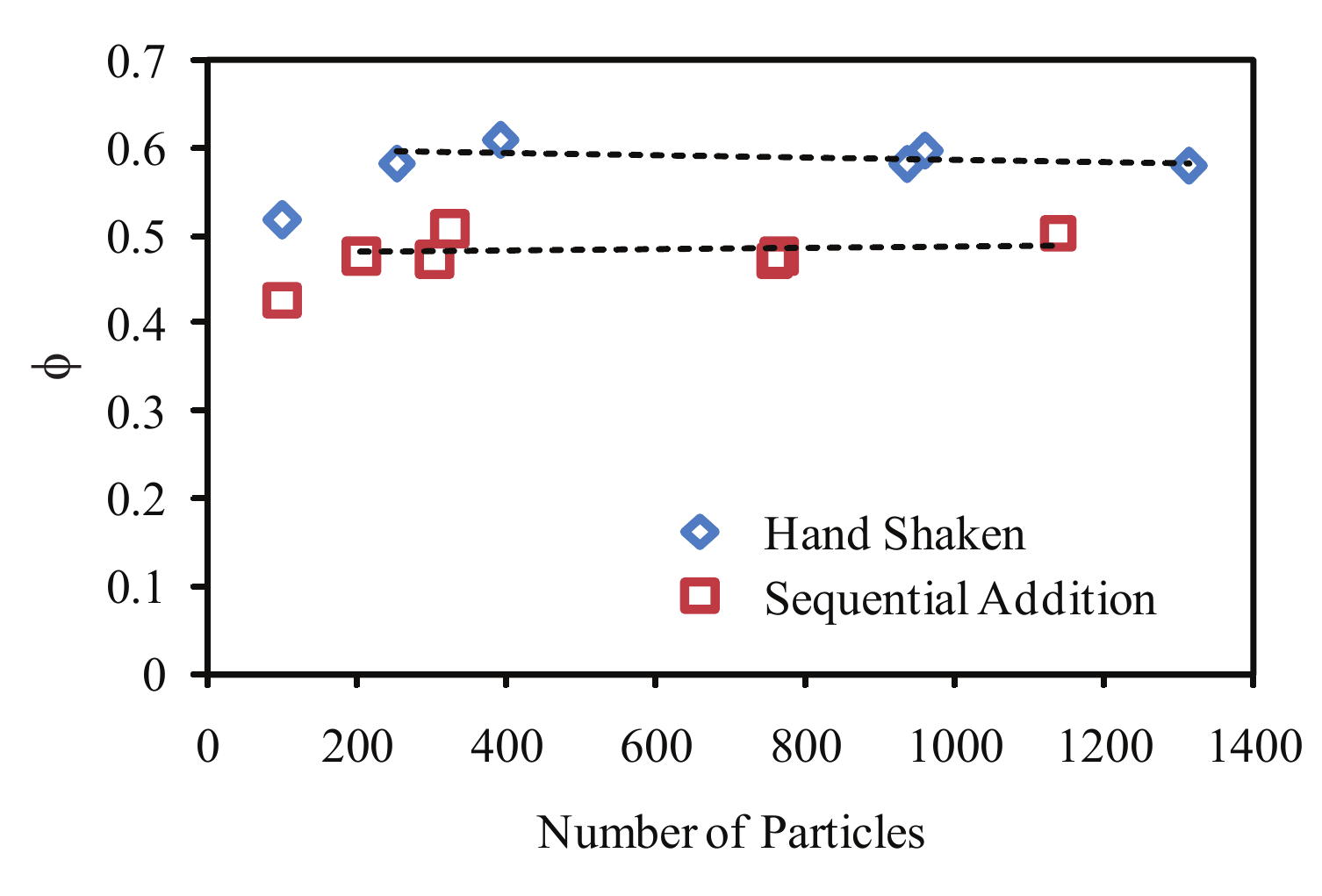}
\caption{(Color online) The volume fraction is not observed to vary significantly beyond 200 as the number of tetrahedrons are increased by an order of magnitude. The lines are a guide to the eye.}
\label{Ndep}
\end{center}
\end{figure}

Because the plastic dice were expensive, the cost became prohibitive to test with a larger number of dice. However, ceramic tetrahedrons used in grinding media were relatively inexpensive and we were able to test the number dependence in this case. In Fig.~\ref{Ndep}, we plot the measured packing fraction for numbers ranging from around 100 to 1200 in proportionately larger containers. Here we used the sequential addition and hand shaken protocol (the particle weight was too high to use our shaker and fluidization experimental setup in this case.) We find that after $N = 200$, $\phi$ remains constant within experimental error and therefore we believe that $\phi$ reported in Table~\ref{exp_phi} are representative of even larger packings and surface effects are small.  

\subsection{Driving frequency dependence}
\label{freqdep}

\begin{figure}
\begin{center}
{\includegraphics[scale = 0.4]{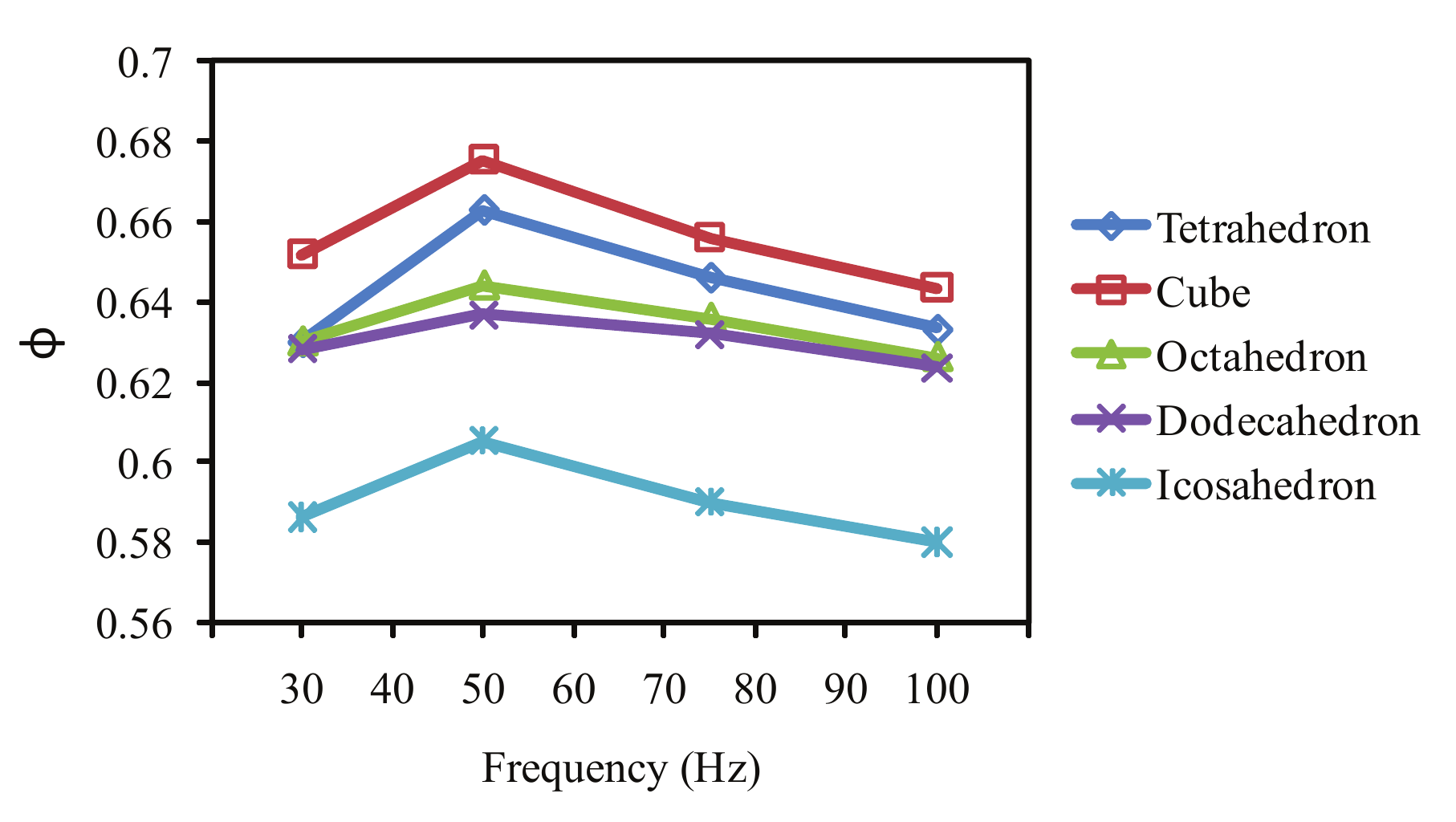}}
\caption{(Color online) The observed packing fraction versus driving frequency. }
\label{shaker}
\end{center}
\end{figure}

We measured $\phi$ for each of the Platonic solids as a function of frequency while keeping the driving strength $\Gamma/f$ constant (see Fig.~\ref{shaker}). A peak is observed at $f = 50$ Hz. The appearance of peak in frequency can be explained as follows. At the lowest frequencies, the particles are tossed up which appears to result in lower packing fractions. Whereas, at high frequency particles do not appear to receive sufficient energy to rearrange from initial packings formed after particles are added. 

\subsection{Characterization of randomness of packing}\label{sec:order}

\begin{figure}
\begin{center}
\includegraphics[width=7cm]{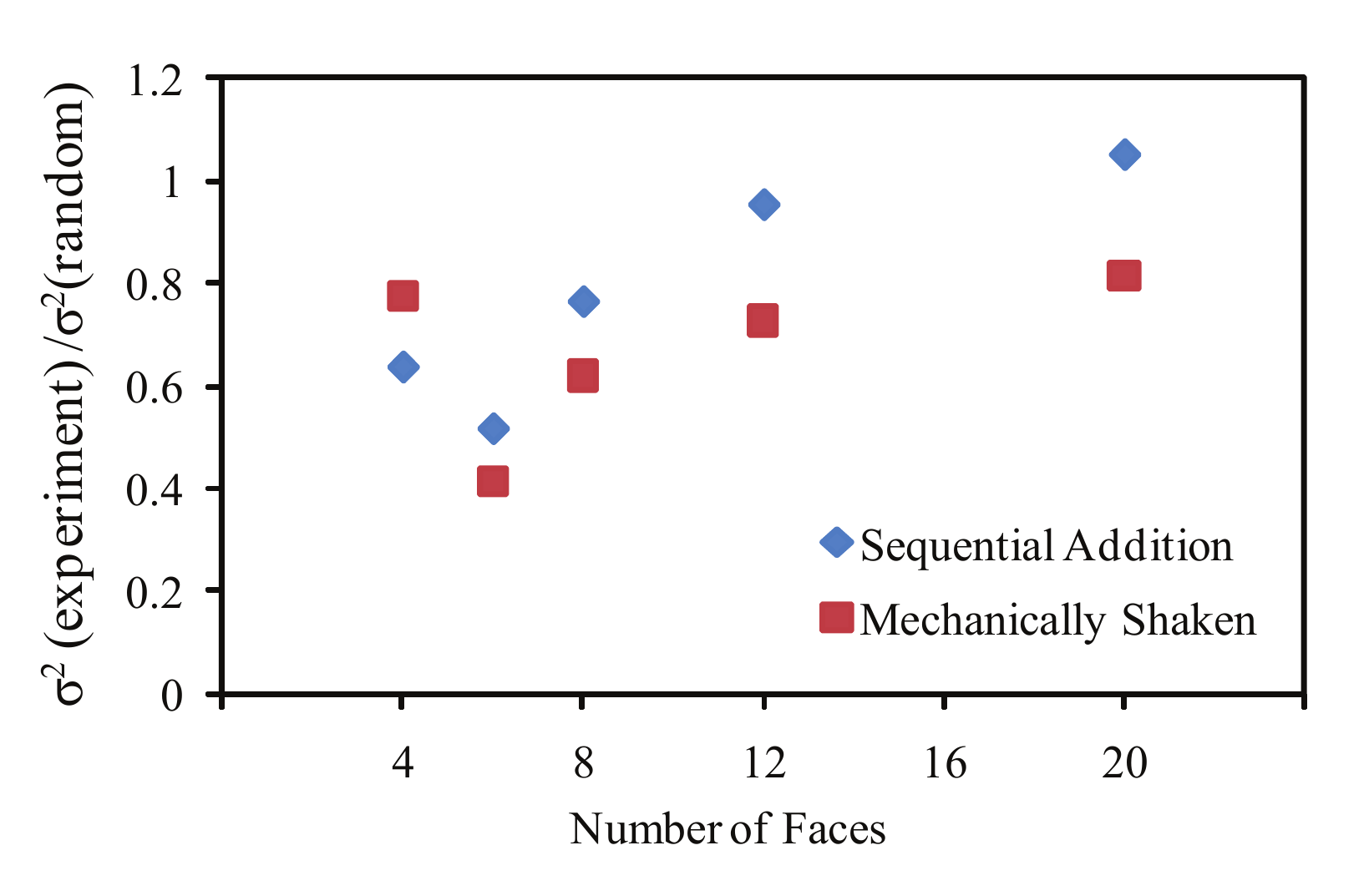}
\caption{(Color online) The ratio of the variance of observed projected areas and that of random projected areas. The data is closer to 1 which indicates random packing rather than to zero, which would correspond to an ordered system.}
\label{order}
\end{center}
\end{figure}

In order to parameterize the randomness in the packings, we use the images of the packing obtained from the top (as shown in Fig.~\ref{topimage}) and measure the projected area of the largest visible face. All the projected areas found lie above a minimum ($0.333, 0.408, 0.577, 0.795, 0.795$ for the 4,6,8,12 and 20 sided polyhedron respectively) which is dependent on the geometry of the polyhedron because the projected area is normalized with the area of the face when laying normal to the perpendicular. Because of the normalization, the maximum projected area is 1. We first sample 100 projected area for each Platonic solid and preparation technique. Then we calculate the variance among these projected areas (see Fig.~\ref{order}). For a periodically ordered system, all the particles would have the same orientation, and therefore the same projected area for the top face. This implies that the variance of these areas would be zero. As a point of comparison, we find the variance of 100 random numbers generated in the allowed range for each shape. This variance represents that of a highly random system. We find the variance of the projected areas for all shapes except the mechanically shaken cube to be closer to the random number variance than it is to the zero variance for an ordered orientation (see Fig. \ref{order}). As cubes have a strong tendency to align, the variance of the projected areas lies slightly below $0.5$ and closer to the ordered extreme for the mechanically shaken case. From this we conclude that in most cases we have disordered or random packings.

\section{Comparison with sphere packing}

In Fig.~\ref{maxpacking}, we also plot the maximum $\phi$, $\phi_{rcp}$ and $\phi_{rlp}$ for spheres along those for the Platonic solids.  The sphere can be considered as a limit of polyhedral shaped particles as the number of sides goes to infinity, and gives a context to understand the observed packing fractions. As has been noted previously, $\phi_{max}$ for all the Platonic solids exceed that for spheres, consistent with Ulam's conjecture that convex particles pack to a greater $\phi$ than spheres. Examining the known maximum packing structure for Platonic solids~\cite{betke,chen10} it appears that faces align near contacts to give the greater $\phi$.

However for dense packings, $\phi_{mrj}$ for spheres lies below only the cube and octahedron, but as the trend decreases from the octahedron to the dodecahedron and icosahedrons, it appears that the sphere value would now be approached from below. Finally, in considering the loose extreme, we find that all the Platonic solids pack less dense than $\phi_{rlp}$ for spheres. This leads us to conjecture that all convex particles in the random loose packed configuration pack less dense than a corresponding packing of spheres.

\section{Summary}

In summary, we have obtained a wide range of packing fractions which depend on the number of faces and edges of the packing solid. The values systematically depend on the protocols used to create these random packings. We hypothesize that the limiting values we report correspond approximately to random loose packing and random dense packing for these shapes. Interestingly, we find the overall trend  similar to that for the maximum packing fraction. In contrast with the maximum packings fraction which always exceed that for spheres, the random packings for Platonic solids with large number of sides pack looser than for spheres. In closing, we note that our packings of idealized polyhedral particles created with random protocols may give better insight into true packings of faceted particles such as sand found in nature.

\begin{acknowledgments}
The work was supported by the National Science Foundation under Grant No. DMR-0605664. We appreciate comments on the manuscript by Salvatore Torquato. 
\end{acknowledgments}

\end{document}